\documentclass[oneside, a4paper, onecolumn, 11pt]{article}
\usepackage[left=2.5cm,top=2.5cm,bottom=2.5cm,right=2.5cm]{geometry}
\usepackage{amsmath}
\usepackage{graphicx}
\usepackage{fancyhdr}
\usepackage{calc}
\usepackage{amssymb}
\usepackage[compress,numbers,square]{natbib}
\usepackage{setspace}
\usepackage{amsfonts}
\usepackage{commath}
\usepackage[innercaption]{sidecap}
\usepackage[framemethod=TikZ]{mdframed}
\usepackage{hyperref}
\usepackage{parskip}
\usepackage{pifont}

\newcommand{\Rom}[1]{\expandafter\@slowromancap\romannumeral #1@}
\makeatother
{ \normalfont} 

\newcommand{\av}[1]{\langle #1 \rangle}

\expandafter\def\expandafter\normalsize\expandafter{%
	\normalsize
	\setlength\abovedisplayskip{0pt}
	\setlength\belowdisplayskip{5pt}
	\setlength\abovedisplayshortskip{0pt}
	\setlength\belowdisplayshortskip{5pt}
}

\definecolor{Gray}{gray}{0.75}

\newmdenv[backgroundcolor=Gray, leftmargin = 0pt, rightmargin = 0pt, linewidth = 0pt, roundcorner = 2 pt, innerleftmargin=5pt, innerrightmargin=5pt, innertopmargin=5pt, innerbottommargin=5pt]{Frame}

\begin{document}

	\newcommand{\kk}{\langle k \rangle}
	\newcommand{\kkk}{\langle k^2 \rangle}
	\newcommand{\er}{Erd\H{o}s-R\'{e}nyi}
	\newcommand{\red}[1]{{\color{red} \bf \footnotesize  #1}}
	\newcommand{\blue}[1]{{\color{blue} #1}}
	\newcommand{\subfigimg}[3][,]{%
		
		\setbox1=\hbox{\includegraphics[#1]{#3}}
		\leavevmode\rlap{\usebox1}
		\rlap{\hspace*{30pt}\raisebox{\dimexpr\ht1-2\baselineskip}{#2}}
		\phantom{\usebox1}
	}
	
	\linespread{1.2}
	
	
	\begin{center}
		{\color{blue} \huge \textbf{Epidemics on evolving 	networks 
		\\[5pt]
		with varying degrees}}
		
		\vspace{2mm}
		Hillel Sanhedrai$^{1*}$ \& Shlomo Havlin$^{1}$
	\end{center}
	
	\small{
		\begin{enumerate}
			\item
			\textit{Department of Physics, Bar-Ilan University, Ramat-Gan, Israel}			
		\end{enumerate}
		
		\begin{itemize}
			\item[\textbf{*}]
			\textbf{Correspondence}:\ \textit{hillel.sanhedrai@gmail.com}
		\end{itemize} 
	}
	
	\vspace{4mm}
	
	\textbf{
		Epidemics on complex networks is a widely investigated topic in the last few years, mainly due to the last pandemic events. Usually, real contact networks are dynamic, hence much effort has been invested in studying epidemics on evolving networks. Here we propose and study a model for evolving networks based on varying degrees, where at each time step a node might get, with probability $r$, a new degree and new neighbors according to a given degree distribution, instead of its former neighbors. We find analytically, using the generating functions framework, the epidemic threshold and the probability for a macroscopic spread of disease depending on the rewiring rate $r$. Our analytical results are supported by numerical simulations. 
		We find surprisingly that the impact of the rewiring rate $r$ has qualitative different trends for networks having different degree distributions.
		That is, in some structures, such as random regular networks the dynamics enhances the epidemic spreading while in others such as scale free the dynamics reduces the spreading. 
		In addition, for scale-free networks, we reveal that fast dynamics of the network, $r=1$, changes the epidemic threshold to nonzero rather than zero found for $r<1$, which is similar to the known case of $r=0$, \textit{i.e.}, a static network. Finally, we find the epidemic threshold also for a general distribution of the recovery time. 
}
	
	\pagenumbering{arabic}


\blue{
\section*{Introduction}
}

Following the fundamental works on epidemic processes \cite{Bernoulli1760, kermack1927contribution, anderson1992infectious, keeling2011modeling},
the study of epidemics on \emph{complex networks} \cite{Albert2002,Newman2010,Havlin2010} has attracted significantly the network science community, and yielded many studies \cite{PastorSatorras2001prl, newman2002spread, Barthelemy2005, keeling2005implications, Dorogovtsev2008, bengtsson2015using, kramer2016spatial,  PastorSatorras2015, Kitsak2010, pei2014searching, van2008virus}, that look mainly for the epidemic threshold for a major outbreak for a variety of static networks and for several epidemic models. 
However, real networks of epidemiological contacts are usually not stationary but show significant dynamic patterns, which challenge the theory for epidemics on static networks \cite{bansal2010dynamic, hamede2009contact, fefferman2007disease, barabasi2005origin}. 

Hence, many studies have investigated epidemic processes on \emph{temporal networks} \cite{gross2006epidemic, volz2007susceptible,volz2009epidemic, valdano2015analytical, leitch2019toward, britton2016network, ball2019stochastic, jiang2019sir, perra2012activity, taylor2012pre,Starnini2014pre,zino2016prl, prakash2010virus}.
Few temporal contact network models have been proposed and studied in the context of epidemic spread.
Volz and Meyers \cite{volz2009epidemic} examined the effect of social mixing on SIR disease.  
They considered \emph{neighbor exchanges}, in which pairs of edges are selected uniformly randomly and swapped, at a fixed mixing rate. Thus, each individual maintains a fixed number of concurrent
contacts while the identities of the contacts change stochastically over time. They found that the epidemic threshold depends also on the rate at which the network changes over time, in addition to the properties of the disease and the network topology. Further studies \cite{britton2016network, ball2019stochastic, jiang2019sir} assumed, rather than an exchange between two random edges, that one side of an edge might be rewired to a random node, or that the edge might be deleted, resulting in non-stationary degrees.
Perra et al.\ \cite{perra2012activity} introduced the \emph{activity-driven network} model, where each node is assigned a time-invariant activity rate. Then, at each time step, each node becomes active with its activity rate, and forms fixed $m$ connections with random nodes. All connections are cleared between time steps. Several works have been done on epidemics on the activity-driven network model and its extensions. \cite{taylor2012pre,Starnini2014pre,zino2016prl}
Prakash et al.\ \cite{prakash2010virus} examined the epidemic threshold under SIS dynamics for arbitrary temporal networks, represented by a sequence of $T$ static network snapshots with
adjacency matrices $A = \{A_1,A_2, . . .A_T\}$. They showed that the epidemic threshold is then
characterized by the maximal eigenvalue of the product of $T$ matrices.

Here, we propose and analyze a model of temporal network based on any degree distribution, where with some rewiring rate, $r$, each node samples new neighbors according to a \emph{new degree} sampled from the given degree distribution. This assumption represents a real scenario where people change, at some points of time, their neighbors and number of neighbors when \textit{e.g.}\ they take part in different events which might have remarkable different sizes and different participants. 
We study on this evolving network model the SIR model. Using an analytical approach based on the generating functions framework, we find the epidemic threshold and the probability of a major outbreak, depending on the rewiring rate $r$, as well as the properties of the network and the infection rate $\beta$.
One of the main questions we address is: does the dynamics in the connections enhance the capability of the epidemic to spread, or the opposite, mitigate it? We find that the answer changes for different structures of the network, which implies two opposite effects of the rewiring on the pandemic as we discuss below.
We derived analytical theory which predicts quantitatively these surprising phenomena.
Another interesting result that we obtain, is about scale-free networks. They are known, in the static case, to be extremely vulnerable to epidemics, such that for any non-zero infection rate the epidemic will spread, that is,  $\beta_c=0$ \cite{cohen2000prl,PastorSatorras2001prl}. However, in the dynamic case, we find that if $r=1$, that is, the network changes fast relative to the infection time, then $\beta_c$ becomes nonzero, namely the evolving network is dramatically more robust against epidemics for this case. 
We also discuss the relation between epidemics and directed percolation \cite{parshani2010dynamic, kinzel1983percolation} which can be regarded as the inspiration for our model. We find that the SIR model on evolving networks can be rigorously mapped to a modification of the directed percolation model. 
Finally, we further explore the case of a general nonuniform recovery time and particularly where there is a recovery rate $\gamma$, instead of taking the recovery time to be a single time step. 
We find for this case the phase diagrams that show the conditions for an epidemic spread depending on the rewiring, infection, and recovery rates, ($r,\beta,\gamma$) and the degree distribution $p_k$.


\blue{
\section*{Model}
}

We consider the stochastic Susceptible-Infectious-Recovered (SIR) model in discrete time. According to this model, if an agent is susceptible and it has a contact with an infectious node at some time step, then it is infected by the infective node with probability $\beta$. Once it is infected, it can infect others at the following time step. For simplicity, we analyze first the case in which an infected node can infect others only at the next time, and then it is recovered and cannot infect nor become infected anymore. Below, we further analyze the more complicated case that an infected node is recovered with probability $\gamma$ at each time step.

\begin{figure}[h]
	\centering
	\includegraphics[width=0.8\textwidth]{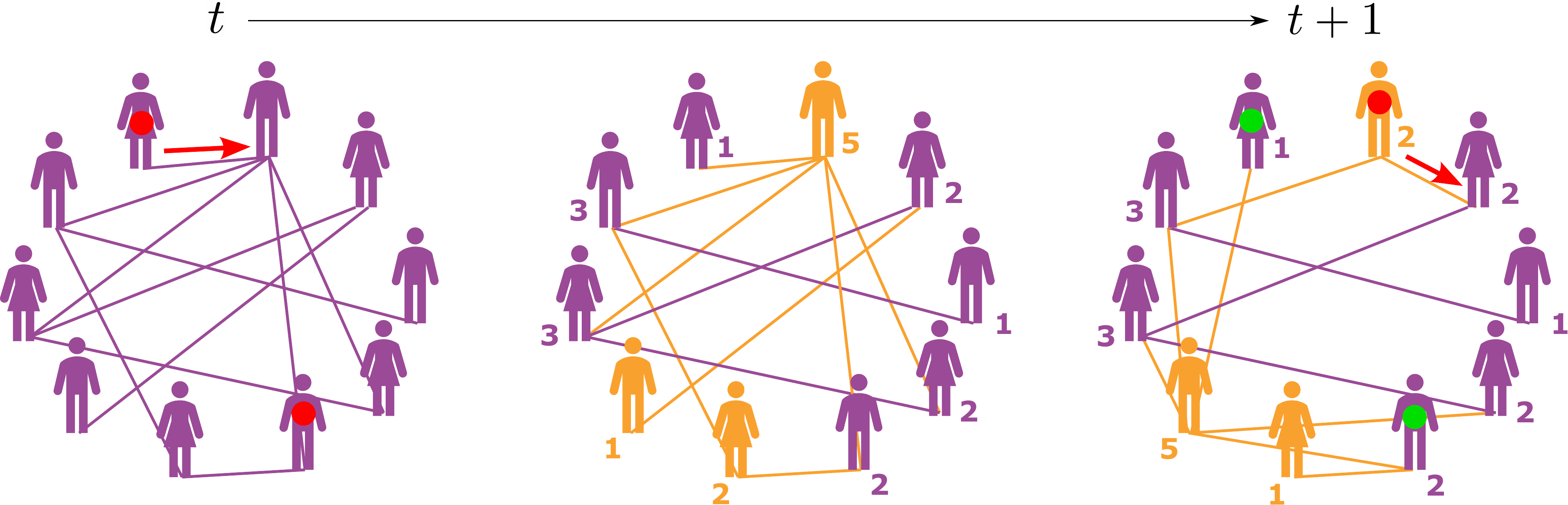}
	\caption{ {\bf Illustration of our evolving network model.}
		When rewiring, first we choose randomly $r$ nodes (in orange) to switch neighbors and degree, while other nodes (purple) conserve their degree. Here $r=0.3$, hence, the outgoing links from the three chosen nodes (orange) are rewired while other links remain. The degrees of the chosen nodes are shuffled randomly, and also their neighbors are shuffled randomly between them. One can see that the degrees of the chosen nodes changed while the degrees of the other nodes were preserved. In such a way, the degree distribution $p_k$ is conserved, and a fraction $r$ of nodes samples new degrees and neighbors. However, notice that also other nodes get some changes in the identities of their neighbors, since if they have a neighbor which is a chosen node, it abandons and leaves someone else instead. The red circles and arrows demonstrate an effect of the rewiring on the pandemic spread. At time $t$ there are two infectious people (red circles). The red arrows represent actual infections. A person with many neighbors is more likely to get infected by one of them. Here comes a major difference between static and dynamic networks. While in the former, the infected node preserves its degree, which tends to be high, in the latter it gets a new degree which is probably smaller. This reduces the epidemic capability to be spread. An opposite effect is that the parent of the infected node is now immunized (green circles) and thus cannot spread the disease, thus switching neighbors has some advantage for spreading. We take all these considerations into account and provide formulas to predict the epidemic spread.
	}	
	\label{fig:illust}
\end{figure}

We present here our model for the evolving network on which the SIR epidemic process takes place. Initially, we build a random network with some degree distribution $p_k$. Then, at each time step, each node is chosen with probability $r$ to switch its degree and neighbors. Next, we shuffle the neighbors of these chosen nodes in the following way. First, we shuffle the degrees of these nodes, hence each node gets randomly a new degree from the pool of degrees of all chosen nodes, thus the degree distribution is conserved. This is since the nodes not selected preserve their degree and the chosen nodes just exchange degrees. Next, we shuffle the neighbors of the chosen nodes, such that each one gets random neighbors according to its new degree from the pool of the neighbors of the chosen nodes. See Fig.\ \ref{fig:illust} for a demonstration of our model for the evolution of the network.

Next, we describe how the epidemic process and the evolution process of the network integrate with each other. 
The infections take place at each time step according to SIR model, whereas the rewiring occurs between time steps. We also assume the recovery occurs between time steps.

 
To summarize, we have the following parameters governing the system behavior. $\beta$ - infecting rate, $r$ - rewiring rate, and for the case that the recovery time is not one, $\gamma$ - recovery rate.

 \newpage


\blue{
\section*{Fixed infectious time $\tau=1$ }
}

For simplicity, we consider first the case in which a node can infect only at the following time after it gets infected, then it gets recovered. That is, the recovery time is $\tau=1$. We look for the probability $P_{\rm inf}$ that a random node infected from outside the system will lead to a macroscopic infection, namely, at the end of the process, an infinite number of nodes have been infected in an infinite network. We further look for the critical infection rate $\beta_c$ for the spread of disease, which represents the transition from zero probability of infinite infection to a nonzero probability.

The probability $P_{\rm inf}$ that a random infected node will lead to an outbreak, that is, spreads the pandemic to an infinite group is the likelihood that it infects at least one of its neighbors and the latter will spread the contagion to infinite. Clearly, it depends on its degree $k$. Thus, the probability, $1-P_{\rm inf}$ , that the infectious node will not spread the epidemic to an infinite number of nodes is,
\begin{equation} \label{eq:Pinf1}
	1-P_{\rm inf} =
	\sum_{k=0}^{\infty}p_k \Big( (1-r)v_{p}^k + r v_{np}^k \Big).
\end{equation}
The RHS means that the infectious node does not spread the epidemic to infinity through anyone of its $k$ neighbors, nor if it preserves its neighbors \emph{after} it infects ($1-r$), 
neither if it switches its neighbors just \emph{after} it infects ($r$). Here, $p_k$ is the degree distribution. The quantity $v_{p}$ is the probability that  the epidemic does not spread to an infinite group \emph{through a random neighbor}, given the parent node was not chosen to switch immediately \emph{after} it had the chance to infect its neighbors. The quantity $v_{np}$ is the same except that the parent node did switch its neighbors the moment \emph{after} it had the chance to infect its neighbors, leaving its children with no parent (np) and with another node instead. The reason we separate this likelihood into two cases, is that otherwise the chances to spread through each neighbor are dependent via the question of whether their parent was chosen to switch or not, since the parent is not susceptible, unlike others.

Defining the generating function \cite{wilf2006,newman2001random} of the degree distribution $p_k$
\begin{equation} \label{eq:G0}
	G_0(x) = \sum_{k=0}^{\infty}p_kx^k , 
\end{equation}
we obtain 
\begin{equation}
	1-P_{\rm inf} = (1-r)G_0(v_{p}) +rG_0(v_{np}) .
	\label{eq:Pinf}
\end{equation}
Note that we do not consider whether the infectious node performed neighbors-switch \emph{before} it infected, since it does not have any impact.

Next, we find $v_p$ and $v_{np}$. Note first that if the parent does not infect its random neighbor (with probability $1-\beta$) then obviously the disease will not spread through this neighbor. Then we analyze what happens if the neighbor got infected. In this case, we distinguish between if it does not switch neighbors (with likelihood $1-r$) \emph{before} it infects or it does. When switching neighbors it gets the degree distribution of a random node $p_k$, while when not switching it preserves its \emph{neighbor} degree distribution which is $kp_k/\av{k}$. We also separate between the case it switches neighbors \emph{after} it infects and the case it does not. Switching $after$ infecting leaves its former neighbors with no parent, consequently gives them the chances $v_{np}$, while not switching gives its neighbors the probability $v_p$ for not spreading the pandemic to infinity. Finally, The difference between $v_{p}$ and $v_{np}$ is where the neighbor did not switch before it infected, then if the parent node was replaced by a random node, there are $k$ susceptible contacts, while if the parent was not replaced, there are $k-1$ susceptible contacts, since the parent is recovered. 
Thus, we obtain two self consistent equations,
\begin{equation} \label{eq:vp}
	v_{p} = 1-\beta + \beta  \left[(1-r)^2 G_1( v_{p}) + (1-r)r G_1( v_{np}) + 
	 r(1-r)G_0( v_{p})+r^2G_0(v_{np})\right] ,
\end{equation}
and
\begin{equation} \label{eq:vnp}
	v_{np} = 1-\beta + \beta \left[(1-r)^2 G_2( v_{p}) + (1-r)r G_2( v_{np}) + 
	r(1-r)G_0( v_{p})+r^2G_0(v_{np})\right] ,
\end{equation}

where $G_1$ and $G_2$, the generating functions of the residual degree and the neighbor degree distributions correspondingly, are defined by
\begin{equation} \label{eq:G1G2}
	\begin{aligned}
		G_1(x) & = \sum_{k=0}^{\infty}\frac{kp_k}{\av{k}}x^{k-1} ,
		\\
		G_2(x) & = \sum_{k=0}^{\infty}\frac{kp_k}{\av{k}}x^{k}	.
	\end{aligned}
\end{equation}
The RHS of Eqs.\ \eqref{eq:vp} and \eqref{eq:vnp} consider in fact four options regarding switching before and after infecting others. For each option, there is the corresponding degree distribution (\emph{before}-switch determines) and the corresponding variable $v_p$ or $v_{np}$ (\emph{after}-switch determines). 
Note that in Eqs.\ \eqref{eq:vp} and \eqref{eq:vnp} we assume that all \emph{random} neighbors (except the parent node) are susceptible because we look on the very first steps of the spread, and at this point, in a large network the probability to catch randomly the recovered or infectious nodes is negligible. 

Notice that the SIS model is different from the SIR model considered here, even at the first steps of the pandemic. The difference is that in SIS the parent node becomes susceptible when its "child" infects others, instead of recovered in SIR. This fact has an impact since there is a non-negligible chance of $1-r^2$ that they are still connected at the time after the parent infected the child. However, for $r=1$ SIS and SIR are equivalent at the very beginning of the process since the parent is not anymore a neighbor of its child after one step and random nodes are susceptible. We further mention that for $r<1$, the SIS model is not solvable by our analysis \cite{parshani2010prl, PastorSatorras2001prl}  since the presence of the susceptible parent makes all its "children" dependent on each other, what prevents us from writing equations such as Eqs.\ \eqref{eq:Pinf}-\eqref{eq:vnp}.

\begin{figure}[h]
	\centering	
	\includegraphics[width=0.99\textwidth]{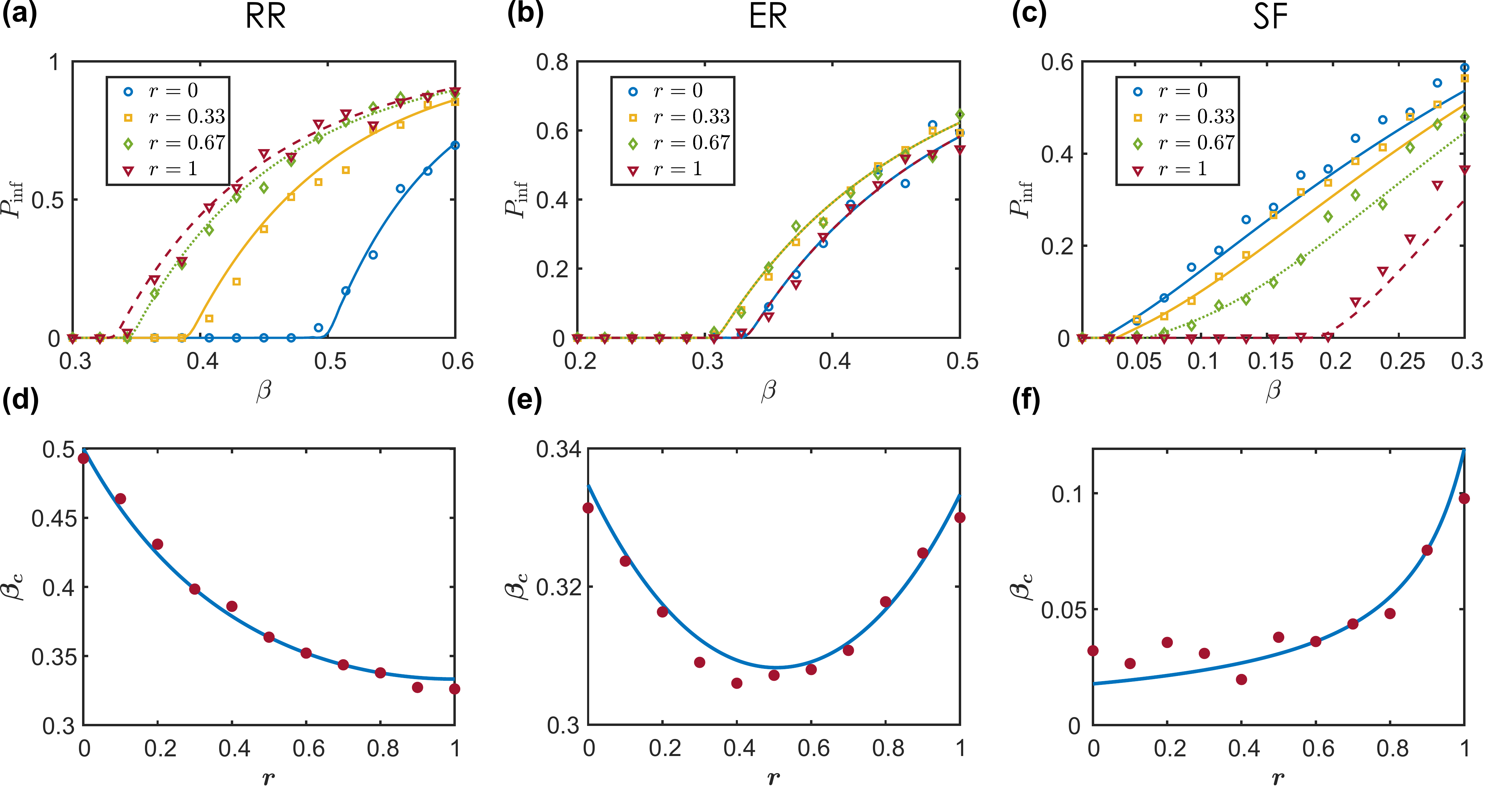}	
	\caption{ {\bf Epidemic spread on evolving networks when the recovery time, $\tau$, is one unit.}
		(a)-(c) The likelihood for large (infinite) infection spreading, $P_{\rm inf}$, versus the infection rate, $\beta$, using Eqs.\ \eqref{eq:Pinf}-\eqref{eq:vnp} (lines) and simulations (symbols). The recovery time is $\tau=1$. In our simulations we set $N=10^4$ and averaged over 300 realizations. The results are for (a) RR network with degree $k=3$, (b) ER network with degree $\av{k}=3$, and (c) SF network with $k_0=2$ and $\lambda=2.5$. In (d)-(f) we show the critical infection rate, $\beta_c$, versus the rewiring rate, $r$, using Eqs.\ \eqref{eq:betac}-\eqref{eq:betacSF} (lines) and simulations (symbols) for (d) RR, (e) ER, and (f) SF. For SF, the simulations agree with Eq.\ \eqref{eq:betac} presented by the line. However, there is some deviation from Eq.\ \eqref{eq:betacSF} due to the finite size effect. See in SI Fig.\ S1 that $\beta_c\to0$ for $N\to\infty$ except at $r=1$.
	}	
	\label{fig:UniformTau}
\end{figure}

Solving Eqs.\ \eqref{eq:vp} and \eqref{eq:vnp} and substituting $v_{p}$ and $v_{np}$ in Eq.\ \eqref{eq:Pinf} we obtain the probability that a random infectious node spreads the epidemic to an infinite group. Fig.\ \ref{fig:UniformTau} shows the analytical results of $P_{\rm inf}$ vs.\ $\beta$ according to Eqs.\ \eqref{eq:Pinf}-\eqref{eq:vnp} compared with simulations, showing excellent agreement. One can see that below $\beta_c$ the chance of infinite infection is negligible, while above $\beta_c$ there is a nonzero probability for it.

\blue{
\subsection*{Critical threshold}
}

Below the criticality ($\beta<\beta_c$), only the single solution $(P_{\rm inf}=0,v_{p}=1,v_{np}=1)$ satisfies Eqs.\ \eqref{eq:Pinf}-\eqref{eq:vnp}, representing no outbreak of the contagion. Above criticality, in contrast, there is another solution representing the existence of a major outbreak, in which $P_{\rm inf}>0$, in addition to the $P_{\rm inf}=0$ solution. 
At criticality, there exists a phase transition between the two states described above, a single solution and a pair of solutions. This determines (see SI Section 1.2) that the derivative of both sides of Eq.\ \eqref{eq:vp} with respect to $v_{p}$ are equal, leading to,
\begin{equation} \label{eq:betac}
	\beta_c =  \frac{1}{(1-r) (\kappa-1+r) + r\av{k}} .
\end{equation}
where $\av{k}=G_0'(1)$ is the average degree, and $\kappa=\av{k^2}/\av{k}=G_2'(1)$ is the average neighbor degree.
The last equation is pretty intuitive because the denominator represents the average number of susceptible contacts which are in touch with an infected node. Thus, it is equivalent to the known condition \cite{anderson1992infectious} of the critical reproductive ratio $R_0=1$.

We can further recognize in this equation that in fact there are two effects of the rewiring. One, the second term in the denominator, $r\av{k}$, is a change of the degree from a typical degree of a neighbor to that of a random node. The second, the adding of $r$ to $\kappa-1$ in the first term, is that the parent node, which infected the current node, might be switched, and replaced by a susceptible node rather than recovered. 

Notice that the limit case $r=0$ recovers the well known result \cite{cohen2000prl,Callaway2000prl,PastorSatorras2001prl} for a static network $\beta_c=1/(\kappa-1)$. The other limit of full rewiring, $r=1$, gives $\beta_c=1/\av{k}$ which recovers the result of branching process since a new degree and neighbors are sampled at each time as in branching process. 

The above results, Eqs.\ \eqref{eq:Pinf1}-\eqref{eq:betac}, are general for any degree distribution. Next, we analyze this result for several model networks. We are interested in the dependence of $\beta_c$ on the rewiring rate $r$. This behavior changes qualitatively for different networks since the relation between $\kappa$ and $\av{k}$ varies.
For random regular network (RR), $\kappa=\av{k}=k$, and substituting this in Eq.\ \eqref{eq:betac} gives, 
\begin{equation} \label{eq:betacRR}
	\beta_c^{\rm RR} =  \frac{1}{ k -(1-r)^2 } ,
\end{equation}
which implies that $\beta_c$ decreases when $r$ increases (see Fig.\ \ref{fig:UniformTau}d). This happens because it is better for mitigating the spread of the disease not to switch neighbors, since when switching, the infectious node is exposed to more susceptible nodes. 
For \er\ network (ER), $\kappa-1=\av{k}$, hence
\begin{equation}
	\beta_c^{\rm ER} =  \frac{1}{\av{k} +r (1-r) } ,
	\label{eq:betacER}
\end{equation}
which interestingly exhibits a non-monotonic behavior (see Fig.\ \ref{fig:UniformTau}e). The reason is that increasing $r$, on one hand, increases the chance that the parent recovered node was replaced by a susceptible node, but on the other hand, it enlarges the probability that the infectious node switches its degree and neighbors, resulting in reduced degree on average.
The most interesting case is scale-free network (SF) whose degree distribution is $p_k=Ak^{-\lambda}$ for $k\geq k_0$ with $\lambda \leq 3$. This distribution has a divergent second moment, and therefore $\kappa\to\infty$. If $\lambda\leq2$ then also the first moment diverges, $\av{k}\to\infty$. Thus, 
\begin{equation} \label{eq:betacSF}
	\beta_c^{\rm SF} = \left\{ 
	\begin{array}{cl}		
		0, & r<1 \text{ or } \lambda \leq2
		\\[5pt]
		\dfrac{1}{\av{k}}, & r=1 \text{ and } \lambda>2
	\end{array}
	\right.	.
\end{equation}
The reason that $\beta_c=0$ for a static SF network is that when spreading the epidemic the infection goes through neighbors. This leads the disease very fast to the hubs, whose degrees are very large, therefore they spread the contagion even if the infection rate $\beta$ is very small. However, in evolving network, when the rewiring rate is one, once an hub gets infected it immediately switches its neighbors and samples a new degree from $p_k$, such that at the moment it comes to infect it is no more a hub. Hence $\beta_c$ becomes nonzero. 
This phenomenon exists in reality when people are likely infected in large events when they have a contact with many potential infectious people, however, only few days later (\textit{e.g.}\ in Covid-19), they can infect, but at this time, they probably do not take part in a mass event.
Interestingly, this finite value of $\beta_c$ is also valid for SIS on evolving networks with $r=1$ as for SIR. This is since for $r=1$, both SIR and SIS are identical at the beginning of the epidemic spreading, see above discussion after Eq.\ \eqref{eq:G1G2}. 
Notice that $\beta_c^{\rm SF}=0$ since $\kappa\to\infty$ for $N\to\infty$. However, for a finite system, $\beta_c^{\rm SF}$ given by Eq.\ \eqref{eq:betac} is finite and dependent of $N$ as shown in Fig.\ \ref{fig:UniformTau}, and in SI Fig.\ S1.

For SF, switching neighbors helps a lot to curb the epidemic, because then the infectious node has a typical degree of a random node which is much lower than the typical degree of neighbors which tend to be hubs. Therefore, when $r$ increases $\beta_c$ also increases in contrast to RR, see Fig.\ \ref{fig:UniformTau}.



\blue{
	\section*{Comparison to directed percolation}
}

It was shown \cite{newman2002spread} that SIR model (with immediate recovery) \emph{on static network} is mapped exactly to \emph{bond percolation}, where the chance $p$ of a link to be occupied replaces the probability $\beta$ to infect. The disease will spread to the whole connected component of the source node. Thus, the chance of a major outbreak equals to the relative size of the giant connected component.

\begin{figure}[h]
	\centering
	\includegraphics[width=0.9\textwidth]{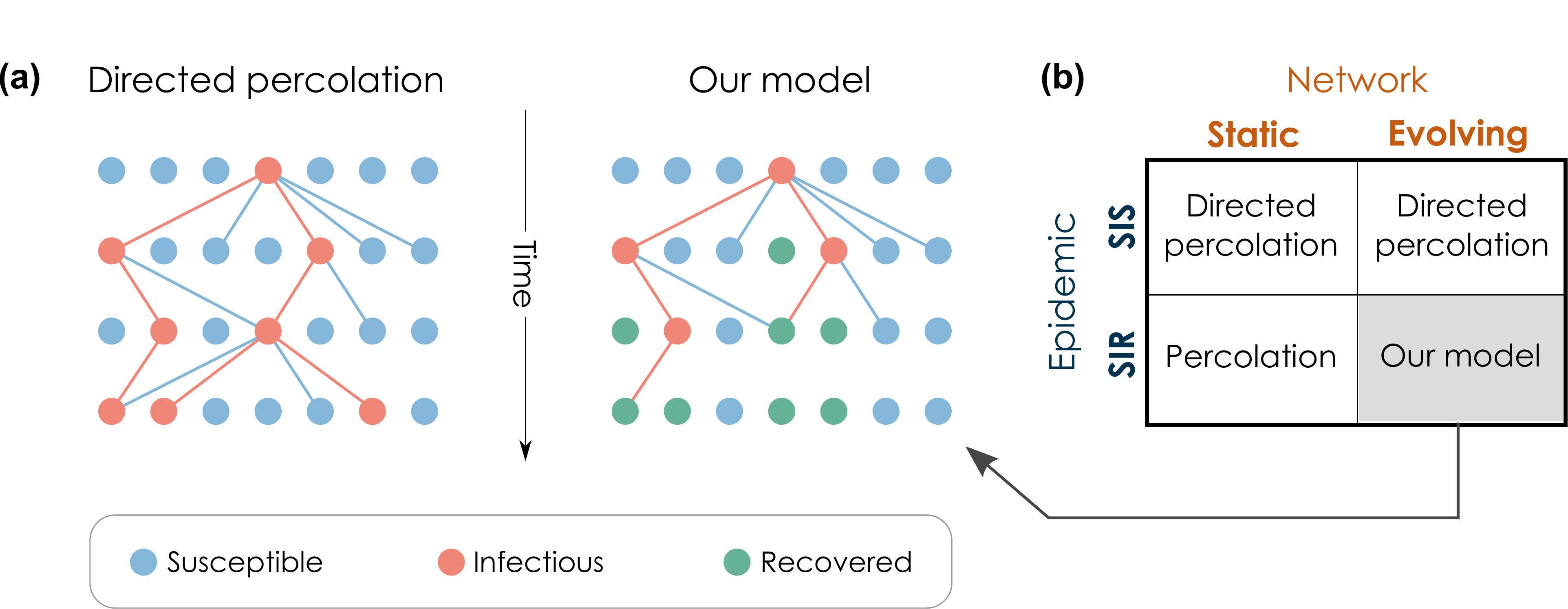}
	\caption{ {\bf Comparison of SIR on evolving networks to directed percolation model.}		
		(a) On the left, we see the directed percolation (DP) model which is equivalent to SIS model. The first row represents the network at time 0, the second at time 1, and so on. The red links represent traverses (in DP) or infections (in SIS), and the blue links indicate contacts that were not traversed (DP) or did not lead to infection (SIS). A link is traversed randomly with likelihood $\beta$. Note that the same node can be infected again and again, allowing the spread of disease. Also the same node might have different degree and neighbors at different times due to the network dynamics. 
		On the right, we show our model (SIR on evolving network) which is identical to the model on left, except that an infectious node recovers after one step and cannot be infected anymore. While on the left (SIS or directed percolation) the epidemic is spread by going through the same nodes few times, on the right the epidemic stops due to the recovery of the spreaders.
		Our model is like directed percolation modified to SIR.
		(b) A table showing the mapping from the four epidemic models on static/dynamic networks to percolation models.
	}	
	\label{fig:illust2}
\end{figure}

SIS model, on the contrary, cannot be mapped to a percolation since a link can be traversed many times. This feature cannot be captured by the single probability $p$ of a link occupation in percolation. However, SIS model can be rigorously mapped to \emph{directed percolation},\cite{kinzel1983percolation} where each time step gets a layer in which a copy of the network \cite{parshani2010dynamic}, see Fig.\ \ref{fig:illust2}. Between the layers there reside the edges of the network at the corresponding time. Each link is traversed with chance $p=\beta$. This mapping covers both static and evolving networks. If the network is static the links between all the layers are identical, while for an evolving network the connections between layers might be different. Of course all links have one direction representing the flow of time.

The only case without mapping to percolation is the SIR model on an evolving network, which can be described nor by percolation since the latter is static, neither by directed percolation in which a node can be traversed many times in contrary to SIR model where an agent can be infected one single time. 

Thus, the SIR model on evolving networks proposed in this manuscript, is actually a modification of directed percolation with the additional following condition. When tracking the cluster reachable from a source, one can go through each node only once. This is since once a node is traversed (infected) it is recovered and immunized  and cannot be traversed again. In our model, the connections from a layer to the next layer are rewired such that each node changes randomly its neighbors and degree with rate $r$. See in Fig.\ \ref{fig:illust2} an illustration of the classic directed percolation compared to our modified directed percolation which represents the SIR model on an evolving network.

\blue{
\section*{Nonuniform infectious time}
}

Next we consider a general distribution for the recovery time $\tau$, and particularly the case in which the probability of an infected node to recover before each time step is $\gamma$. In contrast to the case we analyzed above in which the infectious time was fixed $\tau=1$, now the infectious time is random, distributed exponentially as
\begin{equation} \label{eq:phi}
	\phi(\tau) = (1-\gamma)^{\tau} \gamma .
\end{equation}
Note that also $\tau=0$ is a possible option capturing the scenario of immediate recovery after being infected before infecting others. The mean of this distribution, Eq.\ \eqref{eq:phi}, is $\av{\tau}=(1-\gamma)/\gamma$, ranging from 0 to infinity depending on $\gamma$. The generating function of $\phi(\tau)$ is
\begin{equation}
	G_{\tau}(x) = \sum_{\tau=0}^{\infty} \phi(\tau) x^{\tau} = \frac{\gamma}{1-(1-\gamma)x} .
	\label{eq:Gtau}
\end{equation}
Because of the complexity of this case, we find as above the outbreak probability $P_{\rm inf}$ analytically in the extremes of static ($r=0$) and fast-evolving ($r=1$) networks. For the range in between, we solve only the critical conditions for a major pandemic in the 3D $(r,\gamma,\beta)$ space. 

\newpage

\blue{
\subsection*{Static network \boldmath($r=0$)}
}

An infectious node spends a time $\tau$ in contact with its constant neighbors. The chance to infect each neighbor, $p$, depends on $\tau$ as 
\begin{equation}
	p = 1-(1-\beta)^{\tau}, 
	\label{eq:p}
\end{equation}
which is the complementary probability of no infection in any time during the contact. The infection probabilities of neighbors are dependent on each other through the recovery time of their parent $\tau$, requiring us to separate between different values of $\tau$.
The probability $P_{\rm inf}$ of a major outbreak starting with a random infectious node is
\begin{equation}
	1-P_{\rm inf} =  \sum_{\tau=0}^{\infty} \phi(\tau) G_0(1-p+pv),
	\label{eq:PinfGammaR0}
\end{equation}
where the sum is over all time $\tau$ and requiring that any neighbor does not get infected ($1-p$) or does not spread the pandemic to infinity ($v$) even though it got infected.
The probability $v$ that an infected neighbor does not spread the epidemic to a macroscopic group is obtained by
\begin{equation}
	v = \sum_{\tau=0}^{\infty} \phi(\tau) G_1(1-p+pv).
	\label{eq:vGammaR0}
\end{equation}
Here $G_0$ is replaced by $G_1$ which corresponds to the residual degree distribution (for $v$) rather than the degree distribution (for $P_{\rm inf}$). Eqs.\ \eqref{eq:PinfGammaR0} and \eqref{eq:vGammaR0} are the equivalent of Eqs.\ \eqref{eq:Pinf}-\eqref{eq:vnp} for any recovery time distribution $\phi(\tau)$ and $r=0$.

At criticality, the derivative of both sides of Eq.\ \eqref{eq:vGammaR0} are equal, and substituting Eq.\ \eqref{eq:p}, we obtain (see SI Sec. 2.1) 
\begin{equation}
	\beta_c  = 1 - G_{\tau}^{-1}\left(1- \frac{1}{\kappa-1}\right) ,
\end{equation}
which is valid for any recovery time distribution $\phi(\tau)$. For our case of recovery probability $\gamma$ in every step, using Eq.\ \eqref{eq:Gtau}, we obtain,
\begin{equation}
	\beta_c  = \frac{\gamma}{1-\gamma}\frac{1}{\kappa-2}.
	\label{eq:betacGammaR0}
\end{equation}
The above case of uniform $\tau=1$, is recovered since $G_{\tau}$ is simply the identity function, therefore,
$\beta_c  = 1/(\kappa-1)$.
For the general case of uniform $\tau$, $G_{\tau}(x)=x^{\tau}$, thus $\beta_c=1-[1-1/(\kappa-1)]^{1/\tau}$.
We recognize from Eq.\ \eqref{eq:betacGammaR0} the value of $\gamma_c$ that makes $\beta_c=1$. That is for $\gamma>\gamma_c$ there is no macroscopic outbreak, for any infection rate $\beta$, where
$\gamma_c  = (\kappa-2)/(\kappa-1)$.

\blue{
	\subsection*{Fully temporal network \boldmath($r=1$)}
}

For the case of fully temporal networks, $r=1$, the probability of infecting a neighbor is just $\beta$ independent on $\tau$ since the parent switches its neighbors at each time step. However, the recovery time $\tau$ determines how many neighbors the parent meets before it recovers.
Let $q$ be the number of neighbors that an infectious node meets until it recovers. This number, $q$, satisfies
$q = \sum_{i=1}^{\tau}k_i$,
where $k_i$ are sampled from $p_k$ and $\tau$ is sampled from $\phi(\tau)$. Even if the infectious node is a random \emph{neighbor}, its degree distribution is $p_k$ when it comes to infect, since $r=1$, namely it has new random degree and neighbors. Thus, as a sum of random variables \cite{johnson2005univariate}, $q$ has the average $\av{q} = \av{\tau} \av{k} = (1-\gamma)/\gamma\av{k}$, and using Eq.\ \eqref{eq:Gtau}, its generating function is (see SI Section 2.2)
\begin{equation}
	G_q(x) = G_{\tau} \left( G_0(x) \right) = \frac{\gamma}{1-(1-\gamma)G_0(x)} .
\end{equation}
Due to the fully switching, it does not matter if the spreader is a random node or a random infected neighbor. Hence, $P_{\rm inf} = 1-v$, and
\begin{equation}
	v = G_q(1-\beta + \beta v),
	\label{eq:vGammaR1}
\end{equation}
which yields at criticality,
\begin{equation}
	\beta_c = \frac{1}{\av{\tau}} \frac{1} {\av{k}} =  \frac{\gamma}{1-\gamma} \frac{1} {\av{k}} .
	\label{eq:betacGammaR1}
\end{equation}
The above case of fixed recovery time $\tau=1$ is included by substituting $\av{\tau}=1$ to get $\beta_c=1/\av{k}$.
Here 
$
	\gamma_c  = \av{k}/(\av{k}+1),
$
above which there can not be a major outbreak for any $\beta$. 

\begin{figure}[h]
	\centering
	\includegraphics[width=0.99\textwidth]{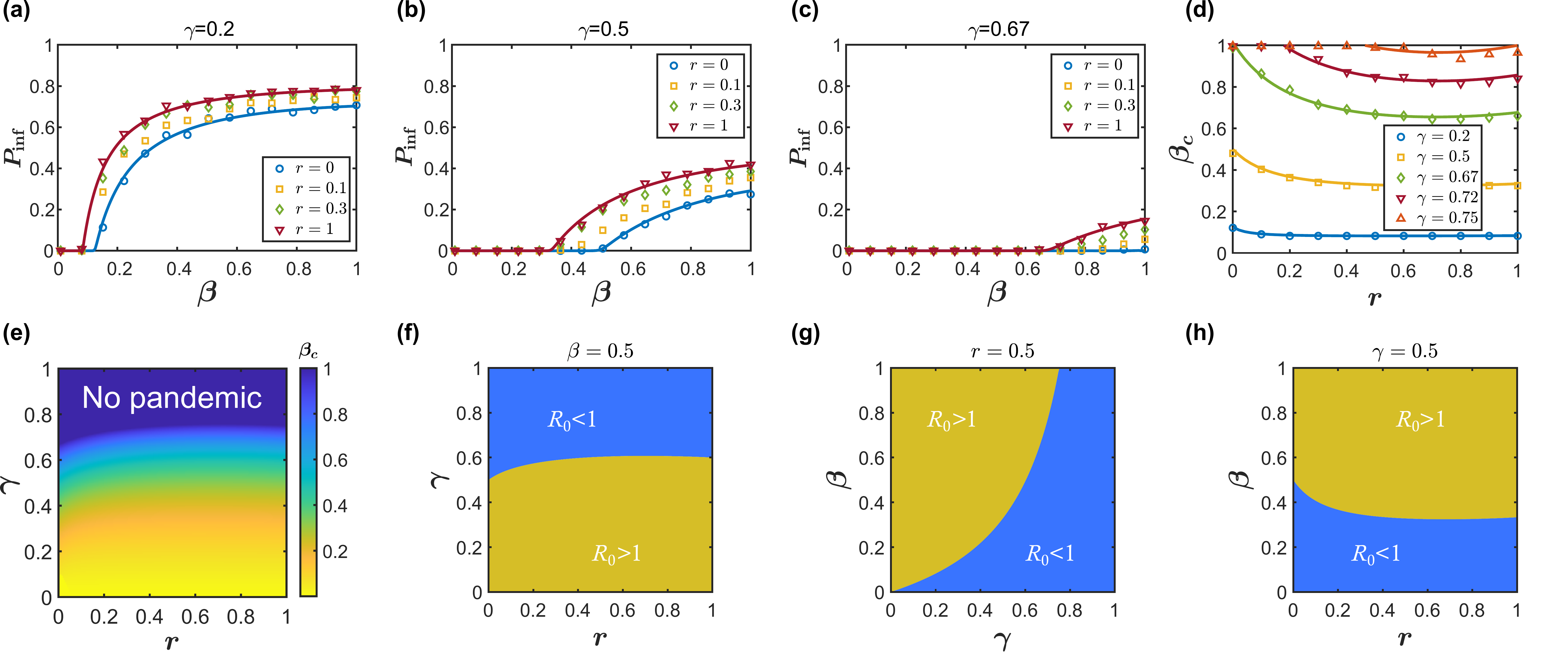}
	
	\caption{ {\bf Epidemic spread on evolving networks for nonuniform recovery time.}
		(a)-(c) The likelihood of a major outbreak, $P_{\rm inf}$, versus the infection rate, $\beta$, for different rewiring rates $r$. Lines represent the theory for $r=0,1$ using Eqs.\ \eqref{eq:PinfGammaR0}, \eqref{eq:vGammaR0} and \eqref{eq:vGammaR1}, and symbols represent simulations results. The recovery rate is presented for values, (a) $\gamma=0.2$ (b) $\gamma=0.5$ and (c) $\gamma=0.67$. In our simulations, we constructed ER networks with average degree $\av{k}=3$ and size of $N=10^4$. We averaged the results over $10^3$ realizations. Since the recovery rate $\gamma=0.5$ yields $\av{\tau}=1$, it is comparable to the above case of fixed $\tau=1$, Fig.\ \ref{fig:UniformTau}b. Note that for $\gamma=0.67$ and $r=0$ there is no transition since $P_{\rm inf}=0$ for any value of $\beta$, see panel (e). 
		(d) The critical infection rate, $\beta_c$ for a major outbreak, versus the rewiring rate, $r$, for different values of $\gamma$ using Eqs.\ \eqref{eq:R0} and \eqref{eq:R01} and simulations. 
		(e)-(f) Eqs.\ \eqref{eq:R0} and \eqref{eq:R01} provide a 3D phase diagram in $(r,\gamma,\beta)$ space which splits into the outbreak phase and the no-outbreak phase. $R_0$ determines the boarders of these phases, $R_0>1$ is the condition for an outbreak. (e) $\beta_c$ below which there is no macroscopic pandemic. Note that in the blue area $\beta_c=1$ \textit{i.e.}\ there is no pandemic for any $\beta$. (f)-(h) Cross-sections in the sub-spaces of $(r,\gamma,\beta)$ for one fixed value (blue - no pandemic, yellow - pandemic).
		In SI Figs.\ S2 and S3, we present the results also for RR and SF networks.
	}	\label{fig:NonuniformTau}
\end{figure}

Comparing Eqs.\ \eqref{eq:betacGammaR0} and \eqref{eq:betacGammaR1} for nonuniform recovery time, we recognize that for ER network where $\kappa=\av{k}+1$, it follows that $\beta_c^{\rm ER}(r=0)>\beta_c^{\rm ER}(r=1)$, implying that the contagion is spread better in fully-evolving network, rather than in a static network. It is interesting that this result is in contrast to the fixed recovery time $\tau=1$, for which $\beta_c^{\rm ER}$ is symmetric for substituting $1-r$ instead of $r$ (Eq.\ \eqref{eq:betacER}). That is, not fixed unit recovery time, causes the dynamics of the network to enhance more the spread of the pandemic. The reason can be understood as follows. Suppose that a node is infectious for a time longer than one unit, then if it switches neighbors it has the opportunity to infect more nodes, while if it stays with the same neighbors it can at most infect all of them, it cannot infect the same node twice. This effect of rewiring does not appear of course at fixed recovery time $\tau=1$. However, Eqs.\ \eqref{eq:R0} and \eqref{eq:R01} imply that also fixed recovery time for any $\tau>1$ and recovery rate $\gamma$ show for ER a specific value of $r$ between $0<r<1$ at which $\beta_c$ is minimal similar to the case of $\tau=1$, see Fig.\ \ref{fig:NonuniformTau}d.

\blue{
	\subsection*{Partial temporal network \boldmath($0<r<1$)}
}

For the partial temporal network, $0<r<1$, we calculate directly the critical conditions for a macroscopic outbreak. To this end, we track the reproductive ratio, $R_0$, defined as the average number of neighbors that a random infected node infects until it recovers. Let us denote by $I$ the random variable of the number of infections that a random infected node performs. $I$ satisfies
\begin{equation}
	I = \sum_{j=1}^{\tau} I_j ,
\end{equation}
where $I_j$ is the number of infections our node acts at time $j$ after it was infected. We seek for $R_0$, which is just 
\begin{equation}
	R_0 = \av{I}.
\end{equation}
To this end, we write a recurrence relation for the degree, $k_j$, and the number of susceptible neighbors, $S_j$, of the infectious node at time $j$ after it got infected (see details in SI Section 2.3), 
\begin{equation} \label{eq:kSj}
	\begin{aligned}
		\av{k_{j}} &= (1-r)\av{k_{j-1}} + r\av{k},
		\\[7pt]
		\av{S_j} & = (1-r)^2 (1-\beta) \av{S_{j-1}} + r(1-r)\av{k_{j-1}} + r\av{k}.
	\end{aligned}	
\end{equation}
for $1<j\leq\tau$.
These equations are based on considering both the possibility that our node switches neighbors just before time $j$, and the possibility that it does not. Even if not, each one of its neighbors might switch and get replaced by another node. We assume as above that any random new neighbor is susceptible since we look at the very first steps of the disease, thus the amount of infectious and recovered nodes is negligible.
Solving this recurrence relation, we manage to obtain an expression for the reproductive ratio (see details in SI Section 2.3),
\begin{equation} \label{eq:R0}
	R_0 = \alpha_1 r \beta	\frac{1-G_{\tau}(\lambda_1)}{1-\lambda_1}+\alpha_2 \beta \frac{1-G_{\tau}(\lambda_2)}{1-\lambda_2} + \av{\tau} \beta \alpha ,
\end{equation}
where $\lambda_1=1-r$, $\lambda_2=(1-r)^2(1-\beta)$, $\alpha = r(2-r)\av{k}/(1-(1-r)^2(1-\beta))$, $\alpha_1 = (1-r)(\kappa-\av{k})/(1-(1-r)(1-\beta))$, $\alpha_2 = y_1-r\alpha_1$, and $y_1 = (1-r)(\kappa-1+r)+r\av{k} - \alpha$.
Thus, $R_0$ captures all the parameters of the problem. It includes the structure-related attributes as well as the epidemic-related characteristics. The structure is represented by $\av{k}$ and $\kappa$ coming from the static pattern of the network and also by the rewiring rate $r$ governing the network dynamics. The epidemic properties are the infection and recovery rates, $\beta$ and $\gamma$.
Next, we use the well-known \cite{anderson1992infectious} critical condition, to get the equation for the critical transition from $P_{\rm inf}=0$ to $P_{\rm inf}>0$,
\begin{equation} \label{eq:R01}
	R_0=1.
\end{equation}
Note that Eqs.\ \eqref{eq:R0} and \eqref{eq:R01} converge, for fixed $\tau=1$, to Eq.\ \eqref{eq:betac} since for this case $G_{\tau}(x)=x$, and $\av{\tau}=1$, thus $R_0=\beta [\alpha_1r+\alpha_2+\alpha ] = \beta [ (1-r)(\kappa-1+r)+r\av{k} ]$. For a general fixed $\tau$, to obtain $R_0$, we just have to substitute in Eq.\ \eqref{eq:R0} $G_{\tau}(x)=x^{\tau}$.


In Fig.\ \ref{fig:NonuniformTau} we show the results for the case of nonuniform recovery time with recovery rate $\gamma$ for ER network (see SI Figs.\ S2 and S3 for RR and SF networks). 
Using Eqs.\ \eqref{eq:PinfGammaR0}, \eqref{eq:vGammaR0} and \eqref{eq:vGammaR1}, we show the probability of a macroscopic spread, $P_{\rm inf}$, depending on the recovery rate $\beta$ for $r=0,1$, showing a good agreement with computer simulations results. For other values of $r$ only computer simulations are presented. Note that larger $\gamma$ gives smaller $\beta_c$ as expected (Fig.\ \ref{fig:NonuniformTau}a-c), and for large enough $\gamma$, and certain values of $r$, there is no macroscopic pandemic spread for any $\beta$ (Fig.\ \ref{fig:NonuniformTau}c). The dependence of $\beta_c$ on $r$, for different $\gamma$ values (Fig.\ \ref{fig:NonuniformTau}d), is different from the dependence for the fixed one unit recovery time, $\tau=1$, shown in Fig.\ \ref{fig:UniformTau}e. While the latter shows non-monotonic and symmetric behavior, the former shows a monotonic increasing pattern except for $\gamma$ close to one. The reason is related, as explained above, to the additional effect of the rewiring when the infectious time is larger than one unit.
For the full range of $0<r<1$, we present (Fig.\ \ref{fig:NonuniformTau}e-f), using Eqs.\ \eqref{eq:R0} and \eqref{eq:R01} the 3D phase diagram in $(r,\gamma,\beta)$ space which splits into the outbreak phase and the no-outbreak phase. This 3D space combines the structure sub-space ($r$) and the epidemic sub-space $(\gamma,\beta)$. $R_0$ is the parameter predicting the borders of these phases, $R_0>1$ leads to an outbreak, while if $R_0<1$ the epidemic will not spread macroscopically. A heat map of $\beta_c$ below which there is no a macroscopic pandemic reveals a region (Fig.\ \ref{fig:NonuniformTau}e, in blue) in which $\beta_c=1$ \textit{i.e.}\ there is no pandemic for any $\beta$. We further present some cross sections of the above mentioned phases in the sub-spaces of $(r,\gamma,\beta)$ where one is fixed (Fig.\ \ref{fig:NonuniformTau}f-g).

%
%
%


\blue{
\section*{Discussion}
}

We explored how an evolving network in which the degree of each node changes randomly according to a fixed degree distribution, responds to epidemics. We found that different network structures show cardinal distinguished behaviors regarding the question: does dynamics in network structure aid the epidemic spread or restrain it? We also found that the well-known character of scale-free networks, the existence of a major outbreak under any infection probability, is broken in a fast enough evolving network, in which the epidemic does not occur for small infection rates. We further found an equation giving the critical condition for a major outbreak depending on all the parameters both of the structure and of the epidemic.

Further study can explore epidemics on evolving networks with some non-homogeneous characters representing better realistic networks. For instance, an evolving network which comprises of few classes of people having different degree distributions. Each person preserves its degree \emph{distribution} during all the epidemic process though its \emph{degree} varies in time. These distributions can share the same shape but differ in the mean or std, or they can even be of completely different shapes. Moreover, the infection rate might be nonuniform among the population \textit{e.g.}\ since part of the people are vaccinated while the rest not as happened in the last waves of the Covid-19 pandemic during 2020-2021. Another interesting direction to explore is the impact of the evolving structure on the efficiency of different vaccinating strategies.


\blue{
	\section*{Acknowledgements}
}
H.S. acknowledges the support of the Presidential Fellowship of Bar-Ilan University,
Israel, and the Mordecai and Monique Katz Graduate Fellowship Program. 
We thank the Israel Science Foundation, the Binational Israel-China Science Foundation (Grant No. 3132/19), the NSF-BSF (Grant No. 2019740), the EU H2020 project RISE (Project No. 821115), the EU H2020 DIT4TRAM, and DTRA (Grant No. HDTRA-1-19-1-0016) for financial support.

\vspace{2cm}

\bibliographystyle{unsrt}
\bibliography{bibliography}

\end{document}


\title{ \color{blue}  \vspace{40mm} \bf {\Huge Epidemics on evolving networks \\[5pt] with varying degrees }\\ \vspace{10mm} Supplementary information}

\maketitle
\thispagestyle{empty}
\clearpage

\tableofcontents
\thispagestyle{empty}
\clearpage
\pagenumbering{arabic}

\blue{
\section{Fixed recovery time $\tau=1$}
}

In this section, we consider the case in which the recovery time is fixed for each infected node, $\tau=1$.

\blue{
\subsection{Major outbreak probability}
}

In Eqs.\ (3)-(5) in the main text, we present the equations for the probability of outbreak in evolving networks,  

\begin{gather}
	1-P_{\rm inf} = (1-r)G_0(v_{p}) +rG_0(v_{np}) ,
	\label{eq:Pinf}
	\\[10pt]
	\label{eq:vp}
	v_{p} = 1-\beta + \beta  \left[(1-r)^2 G_1( v_{p}) + (1-r)r G_1( v_{np}) + 
	r(1-r)G_0( v_{p})+r^2G_0(v_{np})\right] ,
	\\[10pt]
	\label{eq:vnp}
	v_{np} = 1-\beta + \beta \left[(1-r)^2 G_2( v_{p}) + (1-r)r G_2( v_{np}) + 
	r(1-r)G_0( v_{p})+r^2G_0(v_{np})\right] .
\end{gather}

$G_0$, $G_1$ and $G_2$, are the generating functions of the degree, residual degree, and the neighbor degree distributions correspondingly, defined by

\begin{equation}
	\begin{aligned}
		G_0(x) & = \sum_{k=0}^{\infty} p_k x^{k} ,
		\\
		G_1(x) & = \sum_{k=0}^{\infty}\frac{kp_k}{\av{k}}x^{k-1} ,
		\\
		G_2(x) & = \sum_{k=0}^{\infty}\frac{kp_k}{\av{k}}x^{k}	.
	\end{aligned}
\end{equation}
 
Solving Eqs.\ \eqref{eq:vp} and \eqref{eq:vnp} and substituting $v_{p}$ and $v_{np}$ in Eq.\ \eqref{eq:Pinf} we have the probability that a random infectious node spreads the epidemic to an infinite group.

\blue{
\subsection{Critical threshold}
}
Below criticality ($\beta<\beta_c$), only the single solution $(P_{\rm inf}=0,v_{p}=1,v_{np}=1)$ satisfies Eqs.\ \eqref{eq:Pinf}-\eqref{eq:vnp}, representing no outbreak of the contagion. Above criticality, in contrast, there is \emph{another} solution representing the existence of a major outbreak, in which $P_{\rm inf}>0$, in addition to the $P_{\rm inf}=0$ solution. 
At criticality, there takes place a transition between the two states described above, a single solution and a pair of solutions. 
The quantity $v_p$ experiences a transition from a single solution 1 to a pair of solutions 1 and close to 1. At criticality both solutions merge. Since the function of the equation $f(v_p)=0$, determined by Eqs.\ \eqref{eq:vp} and \eqref{eq:vnp}, is analytical, there is a stationary point between the two solutions. When they merge, the stationary point is at 1.   
Thus, if we move $v_p$ in Eq.\ \eqref{eq:vp} to the RHS, the derivative of the RHS with respect to $v_{p}$ is zero. Hence, the derivative of both sides of Eq.\ \eqref{eq:vp} are equal. The same is valid also for Eq.\ \eqref{eq:vnp}, which acts as an equation of the form $f(v_p)=0$ as well. 

The derivatives of Eqs.\ \eqref{eq:vp} and \eqref{eq:vnp} with respect to $v_{p}$ at $v_p=1$ are,

\begin{equation} \label{eq:TwoCritic}
	\begin{cases}				
		1 = \beta_c \Big((1-r)(\kappa-1)+r\av{k}\Big) \Big(1-r + r \dod{v_{np}}{v_{p}}(1) \Big) ,	
		\\[12pt]
		\dod{v_{np}}{v_{p}}(1) = \beta_c \Big((1-r)\kappa+r\av{k}\Big) \Big(1-r + r \dod{v_{np}}{v_{p}}(1) \Big) ,
	\end{cases}
\end{equation}

where $G_0'(1)=\av{k}$ is the average degree, $G_2'(1)=\kappa\equiv\av{k^2}/\av{k}$ is the average neighbor degree, and $G_1'(1)=\kappa-1$ is the average residual degree. As a fundamental character of a probability generating function, it satisfies $G_0(1)=G_1(1)=G_2(1)=1$.
Dividing the second equation in \eqref{eq:TwoCritic} by the first one, we get

\begin{equation*}
		\dod{v_{np}}{v_{p}}(1) =   \frac{(1-r)\kappa+r\av{k}}{(1-r)(\kappa-1)+r\av{k}} .	
\end{equation*}

plugging this into the first equation in \eqref{eq:TwoCritic} gives
\begin{equation*}
1 = \beta_c \Big[(1-r)^2(\kappa-1)+r(1-r)\av{k} + r(1-r)\kappa+r^2\av{k}\Big] ,
\end{equation*}
which finally yields Eq.\ (7) in the main text,

\begin{equation}\label{eq:betac}
	\beta_c = \frac{1}{ (1-r)(\kappa-1+r) + r\av{k} } .
\end{equation}

\blue{
\subsection{Scale-free networks}
}

For scale-free networks (SF) whose degree distribution is $p_k=Ak^{-\lambda}$ for $k\geq k_0$ with $\lambda \leq 3$, the second moment diverges, and therefore $\kappa\to\infty$. If $\lambda\leq2$ then also the first moment diverges, $\av{k}\to\infty$. Thus, from Eq.\ \eqref{eq:betac} we get Eq.\ (10) in the main text,

\begin{equation} \label{eq:betacSF}
	\beta_c^{\rm SF} = \left\{ 
	\begin{array}{cl}		
		0, & r<1 \text{ or } \lambda \leq2
		\\[5pt]
		\dfrac{1}{\av{k}}, & r=1 \text{ and } \lambda>2
	\end{array}
	\right.	.
\end{equation}
 
\begin{figure}[h]
	\centering
	\includegraphics[width=0.5\textwidth]{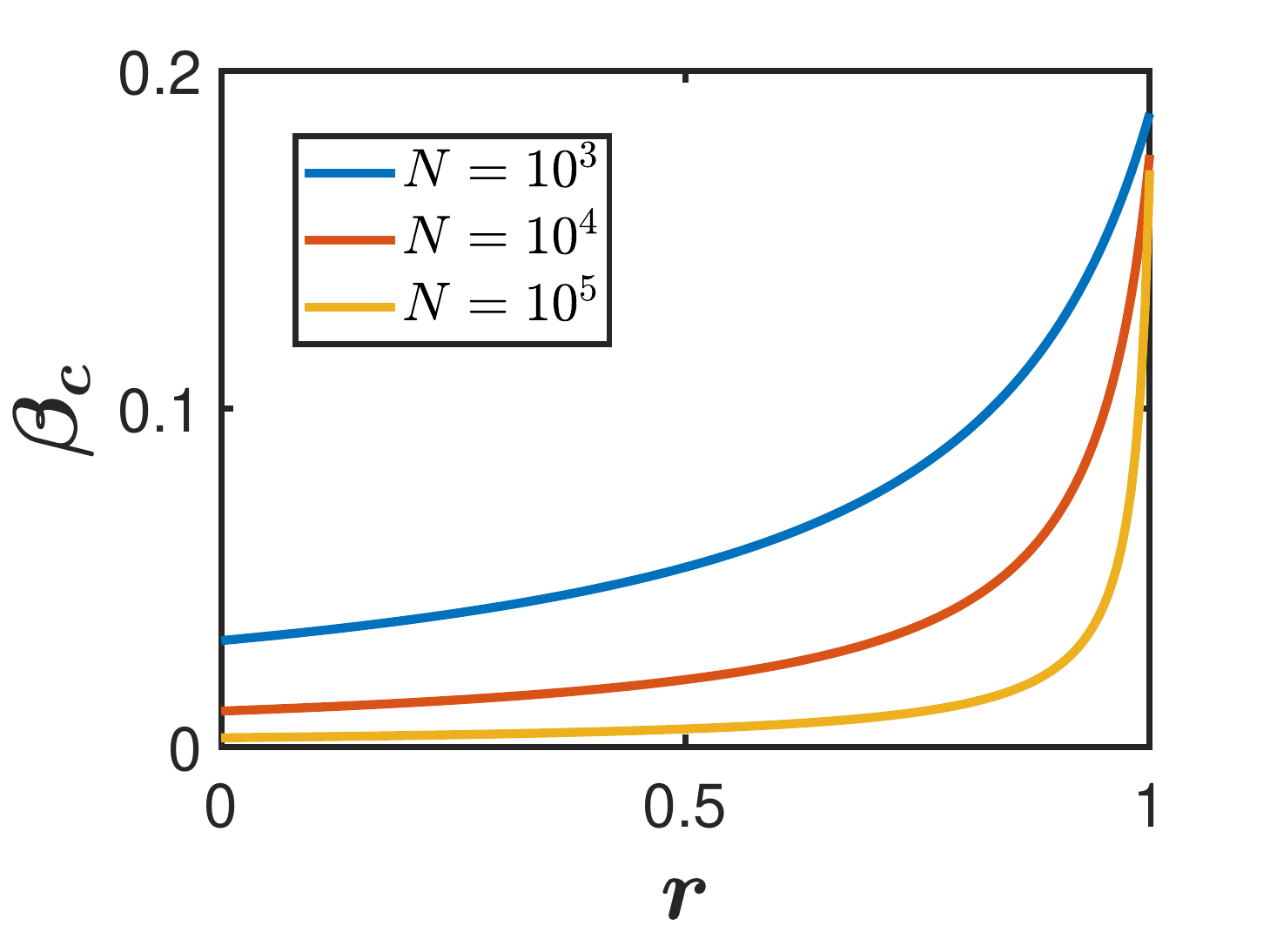}	
	\caption{ {\bf The critical infection rate for evolving SF networks.}
	We show	that when $N$ increases the epidemic threshold, $\beta_c$, approaches zero according to Eq.\ (10) in the main text, except at $r=1$, where $\beta_c$ converges to a non-zero value. This is since for $\lambda=2.5$, $\av{k}$ converges when $N$ increases, while $\kappa\to\infty$. Here we constructed 500 scale free networks with $k_0=2$ and $\lambda=2.5$ to get the mean $\av{k}$ and the mean $\kappa$ of them. We have done this for three sizes, $N=10^3$, $N=10^4$ and $N=10^5$ to.
	}
	\label{fig:betacSF}
\end{figure}

Notice that $\beta_c^{\rm SF}=0$ since $\kappa\to\infty$ for $N\to\infty$. However, for a finite system, $\beta_c$ given by Eq.\ \eqref{eq:betac} is finite and dependent of $N$. However, $\beta_c$ approaches zero as $N$ increases except the singular point $r=1$ at which $\beta_c$ is non-zero, and equals $1/\av{k}$. In Fig.\ \ref{fig:betacSF} we show this for $\lambda=2.5$.

\newpage

\blue{
\section{Nonuniform recovery time}
}

In this section we consider a general distribution for the recovery time $\tau$, and particularly the case in which before each time unit the probability of an infected node to recover is $\gamma$. In contrast to the case we analyzed above in which the infectious time was fixed $\tau=1$, now the infectious time is random, distributed exponentially as

\begin{equation}
	\phi(\tau) = (1-\gamma)^{\tau} \gamma .
\end{equation}

The mean of this distribution is 
\begin{equation} \label{eq:tauMean}
	\begin{aligned}
		\av{\tau} & = \sum_{\tau=0}^{\infty} \tau \phi(\tau) 
		= \sum_{\tau=0}^{\infty} \tau (1-\gamma)^{\tau} \gamma 
		\\
		& = -\gamma (1-\gamma) \frac{\partial}{\partial\gamma}\sum_{\tau=0}^{\infty} (1-\gamma)^{\tau} 
		\\
		& = -\gamma (1-\gamma) \frac{\partial}{\partial\gamma}\frac{1}{1-(1-\gamma)}
		\\[7pt]
		& = \frac{1-\gamma}{\gamma}.
	\end{aligned}	
\end{equation}

The generating function of $\phi(\tau)$ which we use in the main text and below is

\begin{equation}
	G_{\tau}(x) = \sum_{\tau=0}^{\infty} \phi(\tau) =
	\sum_{\tau=0}^{\infty} (1-\gamma)^{\tau} \gamma  x^{\tau} =  \frac{\gamma}{1-(1-\gamma)x} .
	\label{eq:Gtau}
\end{equation}

Because of the complexity of this case, we analyze $P_{\rm inf}$ analytically for the extremes of static ($r=0$) and fast-evolving ($r=1$) networks. For the range $0<r<1$ we find the critical condition for the epidemic spread.

\blue{
\subsection{Static network ($r=0$)}
}

Here we analyze the case of static network, represented by $r=0$.
In this case the chance to infect a neighbor depends on the time of recovery, $\tau$, of the infectious node. We denote this chance by $p$, and find that it depends on $\tau$ of the infectious node as 
\begin{equation}
	p = 1-(1-\beta)^{\tau}.
\end{equation}
Note that the recovery time, $\tau$, of an infectious node makes the chances that it infects each of its neighbors dependent. Hence, the equations in this case are
\begin{equation} \label{eq:vGammaR0}
	v = \sum_{\tau=0}^{\infty} \phi(\tau) G_1(1-p+pv).
\end{equation}
At criticality, this equation experiences a transition from a single solution at 1 to a pair of solutions, 1 and close to 1. Since the function of $f(v)=0$ corresponding to the Eq.\ \eqref{eq:vGammaR0} is analytical there is a stationary point of $f$ between the two solutions. At criticality the stationary point merge with $v=1$. Thus, the derivative at $v=1$ of both sides of Eq.\ \eqref{eq:vGammaR0} are equal. Therefore,
 
\begin{equation*}
	\begin{aligned}
		1 & = \sum_{\tau=0}^{\infty} \phi(\tau) p (\kappa-1).
		\\[4pt]
		1 & = \sum_{\tau=0}^{\infty} \phi(\tau) (1-(1-\beta_c)^{\gamma}) (\kappa-1).
		\\[12pt]
		1 & = (\kappa-1) (1-G_{\tau}(1-\beta_c)) .
	\end{aligned}	
\end{equation*}

which yields

\begin{equation} \label{eq:betacGammaR0}
	\beta_c  = 1 - G_{\tau}^{-1}\left(1- \frac{1}{\kappa-1}\right),
\end{equation}

which is Eq.\ (16) in the main text. \\
For our case of recovery rate $\gamma$, let us find the inverse function of $G_{\tau}$,

\begin{equation*}
	\begin{aligned}
		& y = G_{\tau}(x).
		\\[5pt]
		& y = \frac{\gamma}{1-(1-\gamma)x}.
		\\[5pt]
		& 1-(1-\gamma)x = \frac{\gamma}{y}.
		\\[5pt]
		& x = G_{\tau}^{-1}(y) = \frac{y-\gamma}{(1-\gamma)y}.
	\end{aligned}	
\end{equation*}

Substituting this in Eq.\ \eqref{eq:betacGammaR0}, we get

\begin{equation}
	\begin{aligned}
		\beta_c  & = 1 - \frac{1-1/(\kappa-1)-\gamma}{(1-\gamma)(1-1/(\kappa-1))}
		\\[5pt]
		& = 1 - \frac{(1-\gamma)(\kappa-1)-1}{(1-\gamma)(\kappa-2)}
		\\[5pt]
		& = \frac{(1-\gamma)(\kappa-2)-(1-\gamma)(\kappa-1)+1}{(1-\gamma)(\kappa-2)}
		\\[5pt]
		& = \frac{\gamma}{1-\gamma}\frac{1}{\kappa-2},
	\end{aligned}
\end{equation}

which is Eq.\ (17) in the main text. \\
For uniform $\tau$, the generating function of $\tau$ is $G_{\tau}(x) = x^{\tau}$, thus the inverse function is $G_{\tau}^{-1}(y)=y^{1/\tau}$, which leads to $\beta_c=1-(1-1/(\kappa-1))^{1/\tau}$, covering also the fixed one unit time $\tau=1$, for which $\beta_c=1/(\kappa-1)$.


\blue{
\subsection{Fully temporal network ($r=1$)}
}

In this section all the network is redistributed at each time step. 
Now the probability of infection a neighbor is just $\beta$ independent on $\tau$ since the parent switches its neighbors at each time step. However, the recovery time $\tau$ determines how many neighbors the parent meets before it recovers.
Let $q$ be the number of neighbors that an infectious node meets until it recovers. $q$ satisfies

\begin{equation}
	q = \sum_{i=1}^{\tau}k_i,
\end{equation}

where $k_i$ are sampled from $p_k$ and $\tau$ is sampled from $\phi(\tau)$. Even if the infectious node is a random \emph{neighbor}, its degree distribution is $p_k$ when it comes to infect, since $r=1$, namely it has new random degree and neighbors. Thus, as a sum of random variables\cite{johnson2005univariate}, $q$ has the average

\begin{equation}
	\av{q} = \sum_{\tau=0}^{\infty} \phi(\tau) \av{\sum_{i=1}^{\tau} k_i} = 
	\sum_{\tau=0}^{\infty} \phi(\tau) \tau \av{k} = 
	\av{\tau} \av{k} = \frac{1-\gamma}{\gamma} \av{k} .
\end{equation}

The generating function of $q$ is \cite{johnson2005univariate}

\begin{equation}
	\begin{aligned}
		G_q(x) & = \av{x^q} = \av{x^{\sum_{i=1}^{\tau}k_i}} = \sum_{\tau=0}^{\infty} \phi(\tau) \av{x^{\sum_{i=1}^{\tau}k_i}} 
		\\
		& = \sum_{\tau=0}^{\infty} \phi(\tau) \av{x^k}^{\tau}
		= \sum_{\tau=0}^{\infty} \phi(\tau)\bigg( \sum_{k=0}^{\infty}p_k x^k\bigg)^{\tau}
		\\[10pt]
		& =  G_{\tau} \left( G_0(x) \right).
	\end{aligned}
\end{equation}

Using Eq.\ \eqref{eq:Gtau} it turns to

\begin{equation}
	G_q(x) = \frac{\gamma}{1-(1-\gamma)G_0(x)} ,
\end{equation}

which is Eq.\ (18) in the main text.
Due to the fully switching it does not matter if the spreader is a random node or a random infected neighbor. Hence, $P_{\rm inf} = 1-v$, and
\begin{equation}
	v = G_q(1-\beta + \beta v).
	\label{eq:vGammaR1}
\end{equation}

At criticality, as above in Eq.\ \eqref{eq:vGammaR0}, we equalize the derivative of both sides of Eq.\ \eqref{eq:vGammaR1},

\begin{equation*}
	\begin{aligned}
		& 1 = \beta_c G_q'(1) .
		\\[7pt]
		& 1 = \beta_c \av{q} .
		\\[7pt]
		& 1 = \beta_c \av{\tau} \av{k} .
	\end{aligned}
\end{equation*}

Finally ,we obtain Eq.\ (20) in the main text,

\begin{equation}
	\beta_c = \frac{1}{\av{\tau}} \frac{1} {\av{k}} =  \frac{\gamma}{1-\gamma} \frac{1} {\av{k}}.
\end{equation}


\blue{
\subsection{Partial temporal network ($0<r<1$)}
}

In this interval, we calculate only the critical condition for a macroscopic outbreak. We look on random infected node, and calculate $R_0$, the average number of infections it acts until it recovers. Let us denote by $I$ the number of infections a random infected node performs. $I$ satisfies
 
\begin{equation}
	I = \sum_{j=1}^{\tau} I_j ,
\end{equation}

where $I_j$ is the number of infections our node acts at time $j$ after it was infected. We seek for $R_0$ which is just 

\begin{equation}
	R_0 = \av{I}.
\end{equation}

To find $R_0$, we have to find first $\av{I_j}$. Thus, we track the values of some relevant quantities related to our infectious node along the time until it recovers,

\begin{equation} \label{eq:kSIsj}
	\begin{aligned}
		\av{k_{j}} &= (1-r)\av{k_{j-1}} + r\av{k},
		\\[2pt]
		\av{S_j} & = (1-r) \left( \av{s_{j-1}}+\left(\av{k_{j-1}}-\av{s_{j-1}}\right)r \right) + r\av{k},
		\\[2pt]
		\av{I_j} & = \av{S_j} \beta ,
		\\[9pt]
		\av{s_j} & =\av{ S_j} (1-\beta),
	\end{aligned}	
\end{equation}

for $1<j\leq\tau$.
$j$ captures the time after our infectious node got infected.
$k_j$ is the degree at time $j$. $S_j$ represents the number of susceptible neighbors just \emph{before} the infections of time $j$, $I_j$ is the number of infections done by our node at time $j$, and $s_j$ is the number of susceptible neighbors just \emph{after} the infections at time $j$.
The two first equations in Eq.\ \eqref{eq:kSIsj} are based on the separation for the possibility that our node switches neighbors just before time $j$, and the possibility that it does not. Even if not, each one of its neighbors might switch and get replaced by another node. We assume as above that any random new neighbor is susceptible since we look at the very first steps of the disease, thus the amount of infectious and recovered nodes is microscopic. The last two equations in Eq.\ \eqref{eq:kSIsj} use the infection rate to determine how many nodes, on average, become infectious rather than susceptible. To complete the picture, we add the state at time $j=1$,

\begin{equation} \label{eq:kSIs1}
	\begin{aligned}
		\av{k_1} & = (1-r)\kappa+r\av{k},
		\\[7pt]
		\av{S_1} & = (1-r)(\kappa-1+r) + r\av{k},
		\\[7pt]
		\av{I_1} & = \av{S_1}\beta,
		\\[7pt]
		\av{s_1} & = \av{S_1}(1-\beta).				
	\end{aligned}	
\end{equation}

The reason is that the degree of our infectious node while it gets infected is the typical one of a neighbor, which is on average $\kappa$. The degree after one unit time depends on whether our node performs a switch and gets a random degreewith mean$\av{k}$ or stays with $\kappa$. This explains the first equation. The second equation is different since we assume that all the neighbors are susceptible except the parent node which infected our node. However, there is a chance that the parent switched neighbors meanwhile. Thus, instead of $\kappa$ there is $\kappa-1+r$.
To summarize, we got a recurrence relation for $\av{k_j}$ and $\av{S_j}$,

\begin{equation} \label{eq:kSj}
	\begin{aligned}
		\av{k_{j}} &= (1-r)\av{k_{j-1}} + r\av{k},
		\\[7pt]
		\av{S_j} & = (1-r)^2 (1-\beta) \av{S_{j-1}} + r(1-r)\av{k_{j-1}} + r\av{k}.
	\end{aligned}	
\end{equation}

To solve this, we substitute 

\begin{equation} \label{eq:transform}
	\begin{aligned}	
		\av{k_j} & = x_j+\av{k},
		\\[7pt]
		\av{S_j}& = y_j + \alpha,
	\end{aligned}		
\end{equation}

where

\begin{equation}\label{eq:alpha}
	\alpha = \frac{r(2-r)\av{k}}{1-(1-r)^2(1-\beta)}.
\end{equation}

The new recurrence relation obtained for the new variables $x_j$ and $y_j$ is 

\begin{equation} \label{eq:xyj}
	\begin{aligned}	
		x_j & = (1-r)x_{j-1},
		\\[5pt]
		y_j & = r(1-r)x_{j-1} + (1-r)^2(1-\beta) y_{j-1},
	\end{aligned}		
\end{equation}

and in vectors language,

\begin{equation} \label{eq:vectors}
	\begin{pmatrix}
		x_j 
		\\[4pt]
		y_j 
	\end{pmatrix}
=
	(1-r)
	\begin{pmatrix}
		1 & 0
		\\[4pt]
		r & (1-r)(1-\beta)
	\end{pmatrix}
\begin{pmatrix}
	x_{j-1} 
	\\[4pt]
	y_{j-1} 
\end{pmatrix}
.
\end{equation}

Let us denote this equation by

\begin{equation} \label{eq:uj}
	{u}_{j} = A {u}_{j-1},
\end{equation} 

where $u_j=(x_j,y_j)^T$. The sign $(\ )^T$ represents the transpose operator.
The eigenvalues of the matrix $A$ are 

\begin{equation} \label{eq:lambdas}
	\begin{aligned}
		\lambda_1 & = 1-r,
		\\[5pt]
		\lambda_2 & = (1-r)^2(1-\beta).
	\end{aligned}
\end{equation} 

The corresponding eigenvectors are

\begin{equation} \label{eq:v1v2}
	\begin{aligned}
		v_1&=(1-(1-r)(1-\beta),r)^T,
		\\[5pt]
		v_2&=(0,1)^T.
	\end{aligned}
\end{equation}  

Given the initial conditions $x_1$ and $y_1$, providing ${u}_1=\alpha_1 v_1+\alpha_2 v_2$, we get

\begin{equation}
	{u}_{j} = A^{j-1} {u}_{1} = A^{j-1} (\alpha_1 v_1+\alpha_2 v_2) = \alpha_1 \lambda_1^{j-1} v_1+\alpha_2 \lambda_2^{j-1} v_2.
\end{equation} 

Taking only the second row of the last equation, we obtain

\begin{equation}
	y_{j} = \alpha_1 \lambda_1^{j-1} r+\alpha_2 \lambda_2^{j-1} .
\end{equation} 

Holding this solution for the recurrence relation we move to calculate a quantity leading us towards the desired $R_0$,

\begin{equation}
	I_{\tau} = \sum_{j=1}^{\tau}\av{I_j} ,
\end{equation}

which is the average number of infections that an infected node does \emph{given} it takes time $\tau$ until its recovery. 

\begin{equation}
	\begin{aligned}
		I_{\tau} &= \sum_{j=1}^{\tau}\av{I_{j}} = \sum_{j=1}^{\tau}\av{S_{j}}\beta = \sum_{j=1}^{\tau}\left(y_{j}+\alpha\right) \beta	
		\\[5pt]
		& = \sum_{j=1}^{\tau}\left(\alpha_1 \lambda_1^{j-1} r+\alpha_2 \lambda_2^{j-1} \right)\beta	+ \tau \beta \alpha	
		\\[5pt]
		&= \alpha_1 r \beta	\frac{1-\lambda_1^{\tau}}{1-\lambda_1}+\alpha_2 \beta \frac{1-\lambda_2^{\tau}}{1-\lambda_2} + \tau \beta \alpha .
	\end{aligned}
\end{equation} 

For $R_0$, we average over all possibilities of $\tau$, to get Eq.\ (24) in the main text,

\begin{equation}
	\begin{aligned}
		R_0 &= \av{I} = \sum_{\tau=0}^{\infty} \phi(\tau) I_{\tau} 		
		\\[7pt]
		&= \alpha_1 r \beta	\frac{1-G_{\tau}(\lambda_1)}{1-\lambda_1}+\alpha_2 \beta \frac{1-G_{\tau}(\lambda_2)}{1-\lambda_2} + \av{\tau} \beta \alpha .
	\end{aligned}
\end{equation} 

For criticality we use the known critical condition \cite{anderson1992infectious}

\begin{equation}
	R_0=1.
\end{equation}

$\lambda_1$, $\lambda_2$, and $\alpha$ which appear in $R_0$ are given in Eqs.\ \eqref{eq:lambdas} and \eqref{eq:alpha}.
We still have to find $\alpha_1$ and $\alpha_2$ using the initial condition and the eigenvectors, Eq.\ \eqref{eq:v1v2}. Since ${u}_1=\alpha_1v_1+\alpha_2v_2$, we solve

\begin{equation}
	\begin{pmatrix}
		x_1 
		\\[4pt]
		y_1 
	\end{pmatrix}
	=
	\begin{pmatrix} 
		1-(1-r)(1-\beta) & 0
		\\[4pt]
		r & 1
	\end{pmatrix}
	\begin{pmatrix}
		\alpha_1 
		\\[4pt]
		\alpha_2
	\end{pmatrix}
.
\end{equation}

The solution is 

\begin{equation}
	\begin{aligned}
		\alpha_1 & = \frac{x_1}{1-(1-r)(1-\beta)} ,
		\\[5pt]
		\alpha_2&= y_1 - \frac{rx_1}{1-(1-r)(1-\beta)}.
	\end{aligned}
\end{equation} 

The initial conditions, using Eqs.\ \eqref{eq:kSIs1} and \eqref{eq:transform}, are

\begin{equation}
	\begin{aligned}
		x_1& = (1-r)(\kappa-\av{k})	,
		\\[5pt]
		y_1& = (1-r)(\kappa-1+r) + r\av{k} - \alpha.
	\end{aligned}
\end{equation} 

Therefore, the explicit expressions for $\alpha_{1}$ and $\alpha_2$ are finally

\begin{equation}
	\begin{aligned}
		\alpha_1 & = \frac{(1-r)(\kappa-\av{k})}{1-(1-r)(1-\beta)} ,
		\\[3pt]
		\alpha_2 & = (1-r)(\kappa-1+r)+r\av{k} - \frac{r(2-r)\av{k}}{1-(1-r)^2(1-\beta)} - \frac{r(1-r)(\kappa-\av{k})}{1-(1-r)(1-\beta)}.
	\end{aligned}
\end{equation}

\begin{figure}[h]
	\centering
	\includegraphics[width=0.99\textwidth]{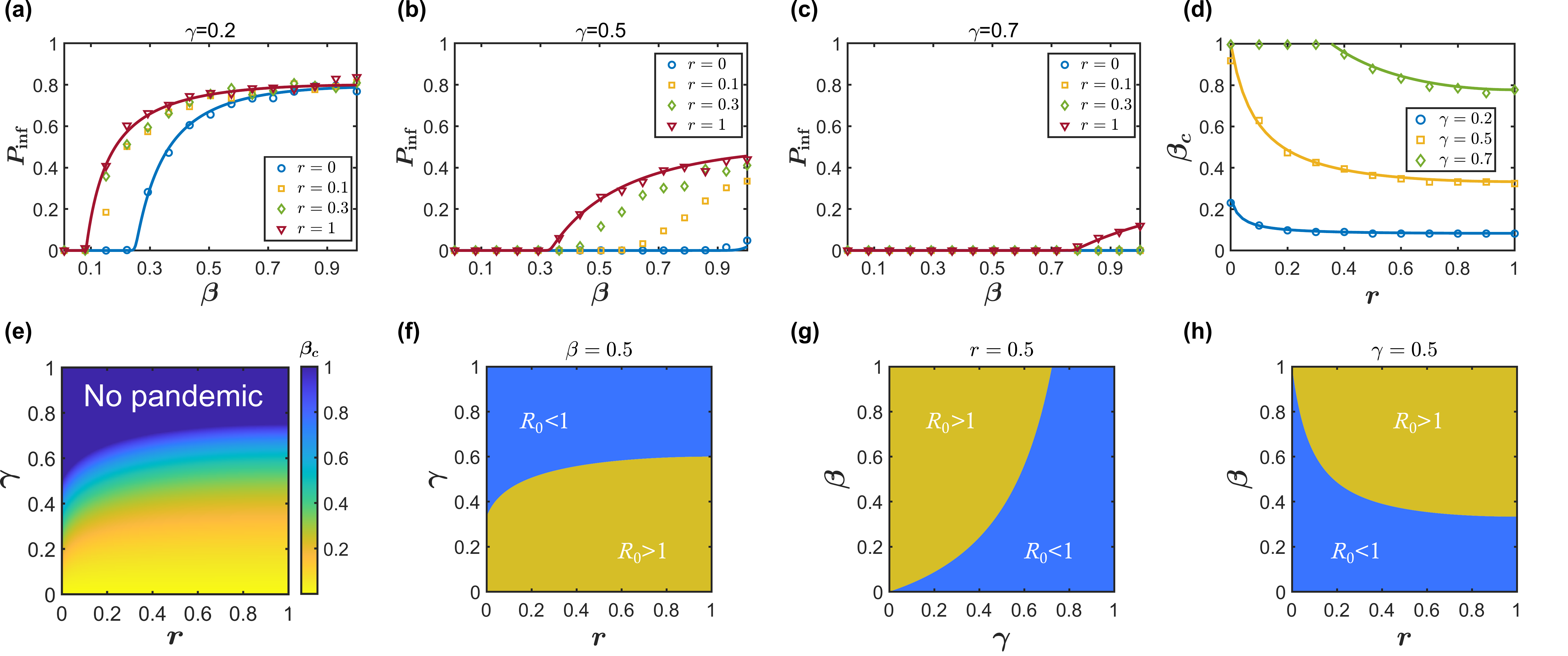}	
	\caption{ {\bf Epidemic spread with recovery rate on evolving RR networks.}
	(a)-(c) The likelihood of a major outbreak, $P_{\rm inf}$, versus the infection rate, $\beta$, for different rewiring rates $r$. Lines represent the theory for $r=0$ and $1$ using Eqs.\ (14), (15) and (19) in the main text, and symbols represent simulations results. The recovery rate was set to be different values, (a) $\gamma=0.2$ (b) $\gamma=0.5$ and (c) $\gamma=0.7$. In our simulations, we constructed RR networks with uniform degree $k=3$ and size of $N=10^4$. We averaged the results over $10^3$ realizations. For $\gamma=0.7$ and $r\neq1$ there is no transition since $P_{\rm inf}=0$ for any $\beta$, see panel (e). 
	(d) The critical infection rate, $\beta_c$ for a major outbreak, versus the rewiring rate, $r$, for different values of $\gamma$ using Eqs.\ (24) and (25) in the main text and simulations. 
	(e)-(f) Eqs.\ (24) and (25) in the main text provide the 3D phase diagram in $(r,\gamma,\beta)$ space which splits into the outbreak phase and the no-outbreak phase. $R_0$ determines the borders of these phases, $R_0>1$ is the condition for an outbreak. (e) $\beta_c$ below which there is no macroscopic pandemic. Note that in the blue area $\beta_c=1$ \textit{i.e.}\ there is no pandemic for any $\beta$. (f)-(h) Slices in the sub-spaces of $(r,\gamma,\beta)$ for one fixed value (blue - no pandemic, yellow - pandemic).
	}	
\label{fig:NonuniformTauRR}
\end{figure}

\begin{figure}[h]
	\centering
	\includegraphics[width=0.99\textwidth]{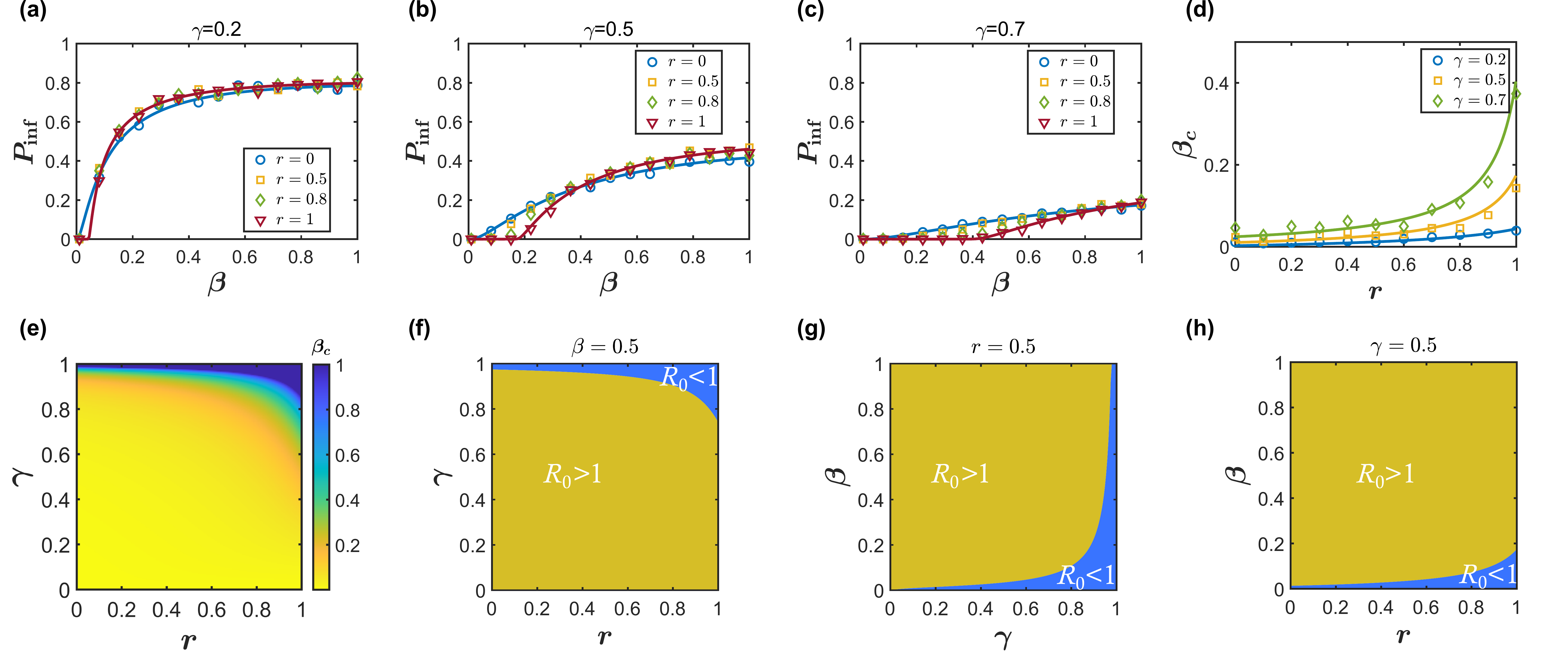}	
	\caption{ {\bf Epidemic spread with recovery rate on evolving SF networks.}
	(a)-(c) The likelihood of a major outbreak, $P_{\rm inf}$, versus the infection rate, $\beta$, for different rewiring rates $r$. Lines represent the theory for $r=0$ and $1$ using Eqs.\ (14), (15) and (19) in the main text, and symbols represent simulations results. The recovery rate was set with varying values, (a) $\gamma=0.2$ (b) $\gamma=0.5$ and (c) $\gamma=0.7$. In our simulations, we constructed SF networks with $k_0=2$, $\lambda=2.5$ and size of $N=10^4$. We averaged the results over $10^3$ realizations.
	(d) The critical infection rate, $\beta_c$ for a major outbreak, versus the rewiring rate, $r$, for different values of $\gamma$ using Eqs.\ (24) and (25) in the main text and simulations. Here we averaged over $500$ realizations.
	(e)-(f) Eqs.\ (24) and (25) in the main text provide the 3D phase diagram in $(r,\gamma,\beta)$ space which splits into the outbreak phase and the no-outbreak phase. $R_0$ determines the borders of these phases, $R_0>1$ is the condition for an outbreak. (e) $\beta_c$ below which there is no macroscopic pandemic. Note that in the blue area $\beta_c=1$ \textit{i.e.}\ there is no pandemic for any $\beta$. (f)-(h) Slices in the sub-spaces of $(r,\gamma,\beta)$ for one fixed value (blue - no pandemic, yellow - pandemic). 
	}	
	\label{fig:NonuniformTauSF}
\end{figure}

In Fig.\ \ref{fig:NonuniformTauRR} we show the results for the case of nonuniform recovery time with recovery rate $\gamma$, for random regular network (RR), and in Fig.\ \ref{fig:NonuniformTauSF} for scale-free network (SF). This is in addition to the results for \er network (ER) presented in Fig.\ 4 in the main text.

\clearpage

\newpage

\blue{
\section{Numerical simulations support}
}

\begin{figure}[h]
	\centering
	\includegraphics[width=0.5\textwidth]{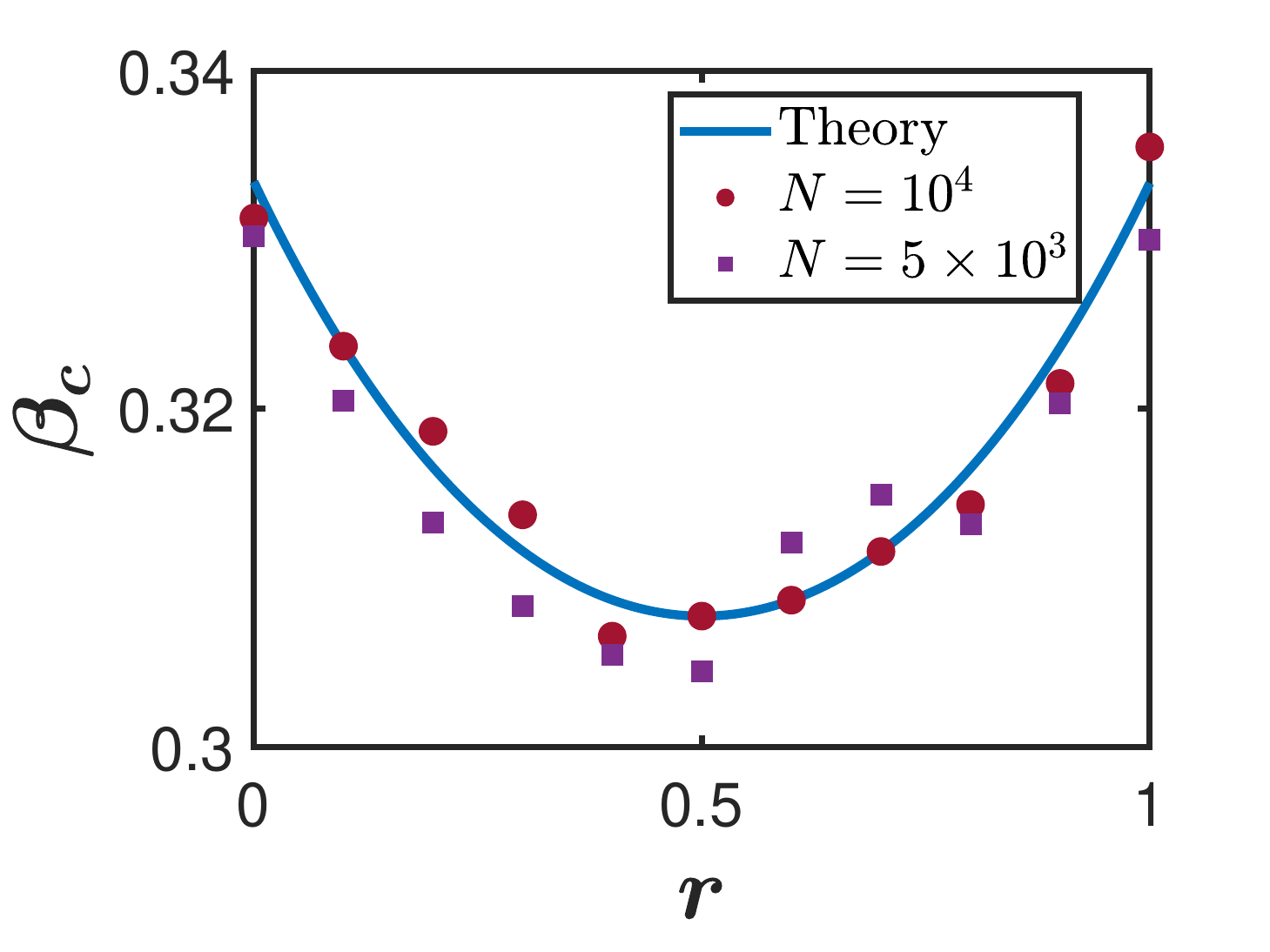}	
	\caption{  {\bf Support of numerical simulations thresholds for different system sizes.} 
	The critical infection rate for an outbreak $\beta_c$ vs the rewiring rate $r$ where the recovery time is $\tau=1$ as shown in Fig.\ 2e in the main text. We show here two system sizes using the same threshold $N/20$ for an "infinite" group, to calculate $P_{\rm inf}$, then we set the same threshold $1.5\times 10^{-2}$ for $P_{\rm inf}$ to find $\beta_c$. We averaged over $10^3$ realizations. We show that the simulations results are close to the theory in both different sizes of the system.  
	}
	\label{fig:ThdiffN}
\end{figure}

To define $P_{\rm inf}$ in the simulations of Figs.\ 2 and 4 in the main text, we had to set a threshold for the number of infections above which the spread is regarded as macroscopic. We set this threshold to be $N/20$. We support this setting in Fig.\ \ref{fig:ThdiffN} where we show similar results for different sizes of system with the same threshold $N/20$.

Similarly, to define $\beta_c$ in the simulations of Figs.\ 2 and 4 in the main text, we had to set another threshold for $P_{\rm inf}$ above which we say that the chance of a major outbreak is non-zero. Since our systems are finite the transition is not very sharp, and thus we should set a threshold larger than zero. We set this threshold to be $10^{-2}$. 
One exceptional case is Fig.\ 2e in the main text where we focus on high resolution thus we tune the threshold for $P_{\rm inf}$ as $1.5\times 10^{-2}$ to obtain a better agreement with the theory. 
The second exception is Fig.\ 3d in SI for SF where the transition is very gradual, hence to catch the criticality accurately we set the threshold for an infinite group as $N/100$, and the threshold of $P_{\rm inf}$ as $2\times 10^{-3}$. In Fig.\ \ref{fig:ThdiffN}, we support the setting of Fig.\ 2e in the main text  where we show similar results for different sizes of system with the same threshold $1.5\times 10^{-2}$.

\bibliographystyle{unsrt}
\bibliography{bibliography}